\documentstyle[osa,manuscript,epsbox]{revtex}  

\begin{document}                

\title{Size Segregation and Convection of Granular Mixtures Almost 
Completely Packed in the Rotating Thin Box.}

\author{Akinori Awazu\footnote{E-mail: awa@zenon.ms.osakafu-u.ac.jp}\\ }

\address{Department of Mathematical Science \\
Osaka Perfecter University, Sakai 593, Japan          } %

\maketitle
\begin{abstract}

Size segregation of granular mixtures which are
almost completely packed in a rotating drum is discussed with 
an effective simulation and a brief analysis. Instead of a 3D drum, we
simulate 2D rotating thin box which is almost completely 
packed with granular mixtures. The phase inversion of 
radially segregated pattern which was found in a 3D experiment are
qualitatively reproduced with this simulation, and a brief 
analysis is followed.
Moreover in our simulation, a global convection appears after
radial segregation pattern is formed, and this convection induces axially 
segregated pattern.\\
\vspace{2mm}
PACS number(s): 45.70.-n, 45.70.Mg\\

\end{abstract}

Granular materials exhibit some complex phenomena\cite{re1}. One
example
is size segregation which occurs by shaking or 
stirring them\cite{s1,s2,s3,s4,s5,s6,s7,s8}.
Mixtures of granular materials which are partially-filled in horizontally 
rotating drum also segregate by size. Recent experiments of such 
rotating drum have shown two types of segregation patterns\cite{s9,s19}.
One of them is radial
segregation where large materials accumulate near the walls of drum and 
small materials accumulate to the central region around the rotating
axis. The other is axial 
segregation where the system evolves to form alternating bands one of 
which is rich in small material and the other is rich in large material. 
These phenomena have been explained by recent numerical 
and analytical studies\cite{s19,s20,s21,s22}. Most of these studies explained
such segregation considering the difference of dynamic angle of reposes
between large and small materials at flowing surface of granular bed.

Recently, radial and axial segregation were observed in
an experiment of a horizontally 
rotating drum which is almost completely packed with granular mixtures\cite{s23}.
Moreover, as explained in the following, phase inversion between two types 
of radial 
segregation patterns takes place when $w$, the angular velocity of the 
rotating drum, passes through a critical value. When $Aw^{2}<g$, large particles
accumulate near the wall of the drum and small particles accumulate to the
central region.
On the contrary, small particles accumulate near the wall of drum and 
large materials accumulate to the inner region of the drum when
$Aw^{2}>g$ \cite{s27}. 
Here, $A$ is the radius of the drum, and $g$ is the acceleration of gravity.
In this system, there are little surface flow because the drum is 
almost completely packed with granular mixtures.
Thus, previous numerical and analytical studies which have taken into
account the difference of dynamic angle of reposes between large and small
materials cannot explain these phenomena. 
In this paper, we discuss the mechanism of such segregation and phase 
inversion phenomena in a rotating drum almost filled with granular 
mixtures. First, we reproduce experimental results with a simple
simulation.
Second, we make an analytical study of the phase inversion phenomena
between two types of radial segregation.

In order to simplify the simulation, we set the following situation. 
Instead of a 3D drum the radius and the width of which are respectively 
$A$ and $B$, we use a 2D box which is rotating 
along a horizontal axis 
with the length $2A$ and the width $B$. Here, the length of 2D box 
corresponds to the diameter of 3D drum. The rotation axis 
of this box is
the line of half length $A$, and this box rotates with angular velocity $w$ 
(Fig.1).
We employ the following particle model which is one of the 
simplest model of granular dynamics\cite{r15}. 
The equation of the motion of the $i$th 
particle is
\begin{equation}
\ddot{{\bf x}_{i}}=-\sum_{j=1}^{N}\theta(r_{i}+r_{j}-|{\bf x}_{i}-{\bf x}_{j}|)\{{\bf \nabla}V(r_{i}+r_{j}-|{\bf x}_{i}-{\bf x}_{j}|)+\eta({\bf v}_{i}-{\bf v}_{j})\}+{\bf F}^{ex}_{i}
\end{equation}
\begin{equation}
V(r_{i}+r_{j}-|{\bf x}_{i}-{\bf x}_{j}|)=\frac{k}{2}(r_{i}+r_{j}-|{\bf x}_{i}-{\bf x}_{j}|)^{2}
\end{equation}
here, $\theta$ is Heaviside function, $N$ is the total number of particles,
$k$ and $\eta$ are respectively the elastic constant and the viscosity 
coefficient, 
and ${\bf x}_{i}(x_{i},y_{i})$, ${\bf v}_{i}(v_{x_{i}}, v_{y{i}})$ and 
$r_{i}$ are, respectively, the position, the velocity and the radius of $i$th 
particles. The elastic constant $k$ and the viscosity coefficient $\eta$ are 
related to the coefficient of restitution $e$ and the collision time 
$t_{col}$, time period during collision\cite{a}. In this model, the effect 
of particles rotation is neglected.
We regard the rotating axis as $x$ axis ($y=0$), and the length direction 
(radius direction for 3D drum) as $y$ direction in this simulation.  
${\bf F}^{ex}_{i}$ is the external force which directly acts on $i$th particle 
not by collision. Since gravitation and centrifugal force work on each 
particle, ${\bf F}^{ex}_{i}(F_{x_{i}}^{ex}, F_{y_{i}}^{ex})$ is given as 
following.
\begin{equation}
F_{x_{i}}^{ex}=0
\end{equation}
\begin{equation}
F_{y_{i}}^{ex}=y_{i}w^{2}-gsin(wt)
\end{equation}
The above equations are calculated with the Euler's scheme. The time step 
$\delta t$ is set enough small such that $\delta x$, the displacement of 
the $i$th particle during $\delta t$, does not exceed a given value.  
We set $2A=7.43$, $B=24.0$, $t_{col}=0.05$, $e$ between a pair of
particles is $0.99$, and $g=3.0$. 
Moreover, boundaries of box $|x|=A$ or $|y|=B$ are given as the 
visco-elastic walls with $e=0.95$.
Total number of particles is $N=750$, the ratio of particle numbers 
between large and small particles is 1:4, the ratio of average radius 
between them is 2:1, and 10\% polydispersity for large and small 
particles' radius is given. The packing density which is defined as [the area of region occupied by particles]/[the area of 2D box]
is estimated 
about 84\%. It means this system is not completely packed. In practice,
the movement of the center of the mass of particles 
appears through the rotation process in our simulation. However, the 
distance of this movement from the average position is almost same as the
average radius of small particles which is enough small compared to $A$. 
Hence, this system is regarded as an almost completely packed system.
Because of such polydispersity and the movement of the center of 
mass of particles, each particle in the system barely moves.

We pack particles at random at the initial condition and simulate with 
several values of $w$.
Figure 2 (a) and (b) are typical patterns of radial segregation for 
(a)$w=0.5$ ($Aw^{2}<g$), and (b)$w=1.5$ ($Aw^{2}>g$). Figure.2 (a) 
indicate that large particles accumulate near the wall of drum ($|y|=A$) 
and small particles gather around the central region when $Aw^{2}<g$. 
Figure.2 (b) indicate small particles accumulate near the wall of drum and 
positions of large particles accumulate to the central region when 
$Aw^{2}>g$. These results qualitatively correspond to the experimental 
results of rotating drum which is almost completely packed with granular 
mixtures\cite{s27}.
The pattern illustrated in Fig.2 (b) is stable  
whereas the pattern like Fig.2 (a) evolves to the pattern like Fig.3
(c). This is because the fluctuation of large
particles concentration in axial direction grows up slowly like followings.
Now, we consider the case that the direction of gravity is the negative 
direction in $y$. It means that  $y=-A$ corresponds to the bottom of 
box. Near the bottom of box, the region in which large particles are
packed exists. Above this region, the region in which small
particle is rich exists. In convenience, we name boundaries between
these two regions S-L-boundary. Because of the fluctuation of large particles
concentration near the bottom, S-L-boundary has finite inclination.
Small particles on this slope cannot invade to the region in which large
particle is rich  because large particles are
packed densely. However, along this boundary, small particles can flow
down. Then, the amount of the small particles flow in $y<0$ region
is very small compared to that in the $y>0$ region because particles in $y<0$
region are packed more densely than in $y>0$ region. 
Moreover, above the region rich in small
particles, the region rich in large particle exists again. We name
the boundary between these two regions L-S-boundary. Large particles in
this region cannot invade into the region in which small particle is
rich because they cannot go through small voids which appear in the
region rich in small particles. Because of friction working at L-S-boundary,
however, some parts of large particles in this region near
L-S-boundary are dragged by the flow of small particles,
which flow down along the slope of
S-L-boundary. Thus, some of large particles around L-S-boundary flow
down along the boundary.
Hence, the axial concentration of large particles in this
region fluctuates and this fluctuation grows up along S-L-boundary
(Fig.3 (a)$\to$(b)$\to$(c)).  
The direction of gravity periodically changes between the
negative direction and the positive direction in $y$ with the rotation 
of the box. Thus,
above mentioned flow of
particles in all around the system
forms global convection 
along S-L-boundaries like Fig.4. Furthermore, the convection makes the 
fluctuation of large particles concentration grow up, and changes the
segregation pattern from the radial to the axial.
Moreover, the axial segregation pattern is kept stable by the convection.

Now, we discuss the mechanism of the phase inversion of the radial 
segregation pattern. In order to discuss this phenomenon, 
we consider following assumptions 
hold according to previous studies\cite{s1,s2,s3,s4,s5,s6,s7,s8}; 
Compared to large particles, small particles can move more
 easily in granular bed because they can move through smaller voids.
It means that small particles are more directly drifted by external
force than large particles. 
Then, we assume that small particles, compared to large particles,
tend to move to the direction in which external force works. Large
particles can also move through voids if sizes of them are larger than
that of large particles.
Then, large particles tend to move to the low particles density region 
in which such large voids tend to appear.
By use of these assumptions, the mechanism of
the phase inversion of radial segregation pattern is discussed. The
force which works on each particle is given by eq.(3) and eq.(4). 
We put $(X,Y)=(y cos(wt),y sin(wt))$. 
Here, the direction of gravitation is from $Y>0$ to $Y<0$. 
Now, the region in drum is given by circle $X^{2}+Y^{2} \le A^{2}$, and 
the region in which $F^{ex}_{i}<0$ for $y>0$ and $F^{ex}_{i}>0$ for $y<0$ are
satisfied is given by the circle 
\begin{equation}
X^{2}+(Y-\frac{g}{2w^{2}})^{2}<(\frac{g}{2w^{2}})^{2}.
\end{equation}
In convenience, we name the former circle 'circle 1' and the latter
'circle 2'. The external forces work toward the center of
circle 1 on each particle in circle 2, and toward the circumference of
circle 1 in another region in circle 1. Now, the movement of particles
in the $Y<0$ 
region seems quite small compared to that in the $Y>0$ region because 
the packing of particles in the $Y<0$ region is more dense than that 
in the $Y>0$ region. 
Hence, we need to consider the movement of particles only in the $Y>0$ region. 
When $Aw^{2}<g$, circle 2 covers up to the circumference of circle 1 in $Y>0$ (Fig.5).
It means that in most of $Y>0$ region, the external force works toward 
the center of drum, so that small particles move toward the 
center of drum. Because of the movement of small particles, low
density region appears near the upper most position of drum
and large particles accumulate around there.
On the contrary, when 
$Aw^{2}>g$, circle 1 completely covers circle 2 (Fig.5). It means that 
small particles in circle 2 move toward the center of drum, and small 
particles out of circle 2 move toward the wall of drum. Then, since low 
particle density space appears around the circumference of circle 2, large 
particles accumulate there. The length between the center of 
circle 1 and the most far point on the circumference of circle 2 from 
the center of circle 1 is given by $\frac{g}{w^{2}}$. Hence, When 
$Aw^{2}>g$, annular pattern the radius of which is estimated about
$\frac{g}{w^{2}}$ is formed by large particles.
Thus, the relation between the 
angular velocity of drum $w$ and the radius $\bar{y}$ where large 
particles aggregate is given by 
\begin{equation}
\bar{y}=A \cdots (w<(\frac{g}{A})^{\frac{1}{2}})
\end{equation}
\begin{equation}
\bar{y} \sim \frac{g}{w^{2}} \cdots (w>(\frac{g}{A})^{\frac{1}{2}}).
\end{equation}  
The phase inversion angular velocity $w_{c}$ is given by 
$w_{c}=(\frac{g}{A})^{\frac{1}{2}}$.

In this paper, by use of simulations and brief analysis, we discussed
the radial segregation and the axial segregation of granular mixtures
which are almost completely packed in a horizontally rotating drum.
By simulating a 2D horizontally rotating box, we reproduced two types of 
radial segregation patterns and the phase inversion between them,
which have been found in previous experiments\cite{s27}. 
Furthermore, in this simulation, global convection was observed to appear
after the radial segregation pattern is formed, 
and this convection caused the axial segregation pattern.
Moreover, by use of the competition between
gravitation and centrifugal force which depends on the angular 
velocity of drum, we explained the phase inversion and found the critical 
angular velocity at which the phase inversion takes place. In order to explain the phase inversion, we assumed
that small particles, relatively, tend to move in the direction of
external force and large particles move to the lower particle density
region. The justification of this assumption remains to be made.

The author is grateful to H.Nishimori, M.Nakagawa and C.Cunkui for useful
discussions. This research was supported in part by the Ibaraki
University SVBL and Grant-in-Aid for JSPS Felows 10376.

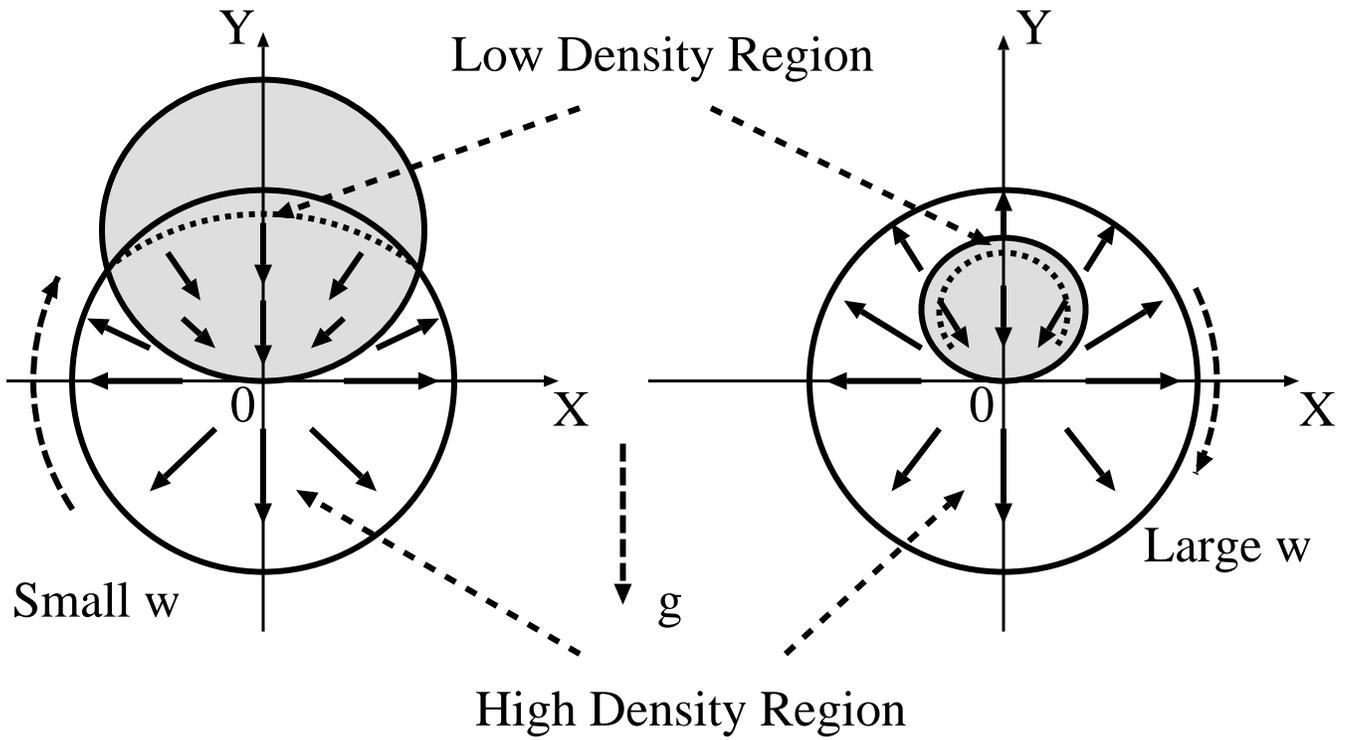
\end{document}